\documentclass[aps,pra,twocolumn,amssymb,amsfonts,floatfix,superscriptaddress]{revtex4}

\usepackage{graphicx}
\usepackage{dcolumn}
\usepackage{bm}
\usepackage{mathbbol}

\usepackage{amsmath,amssymb,amsthm}
\usepackage{times}
\usepackage{graphicx}
\usepackage{bm}
\usepackage{color}
\usepackage{multi row}

\begin{document}

\title{Generic dynamical features of quenched interacting quantum systems: Survival probability, density imbalance, and out-of-time-ordered correlator}

\author{E. J. Torres-Herrera}
\affiliation{Instituto de F{\'i}sica, Benem\'erita Universidad Aut\'onoma de Puebla, Apartado Postal J-48, Puebla, Puebla 72570, Mexico}
\author{Antonio M. Garc\'ia-Garc\'ia}
\affiliation{Shanghai Center for Complex Physics, Department of Physics and Astronomy,
Shanghai Jiao Tong University, Shanghai 200240, China.}
\author{Lea F. Santos}
\affiliation{Department of Physics, Yeshiva University, New York, New York 10016, USA}

\begin{abstract}
We study numerically and analytically the quench dynamics of isolated many-body quantum systems. Using full random matrices from the Gaussian orthogonal ensemble, we obtain analytical expressions for the evolution of the survival probability, density imbalance, and out-of-time-ordered correlator. They are compared with numerical results for a one-dimensional disordered model with two-body interactions and shown to bound the decay rate of this realistic system. Power-law decays are seen at intermediate times and dips below the infinite time averages (correlation holes) occur at long times for all three quantities when the system exhibits level repulsion. The fact that these features are shared by both the random matrix and the realistic disordered model indicates that they are generic to nonintegrable interacting quantum systems out of equilibrium. Assisted by the random matrix analytical results, we propose expressions that describe extremely well the dynamics of the realistic chaotic system at different time scales.
\end{abstract}

\maketitle
{\em Introduction.} Nonequilibrium dynamics of isolated many-body quantum systems is a highly interdisciplinary subject covering a broad range of physics scales, from string theory and black holes to condensed matter and atomic physics. The connection between black hole physics and unitary quantum dynamics emerges from holographic dualities~\cite{Maldacena1998}.  On the experimental side, unitary quantum dynamics is investigated with cold atoms~\cite{kinoshita06,Schreiber2015,Bordia2017,Kaufman2016}, ion traps~\cite{Jurcevic2014,Smith2015}, and nuclear magnetic resonance platforms~\cite{GarttnerARXIV,WeiARXIV}.

Driven by different purposes, studies of black hole information loss~\cite{Roberts2015,Papadodimas2015,CotlerARXIV}, quantum chaos~\cite{Roberts2015PRL,Maldacena2016}, thermalization in isolated quantum systems~\cite{kinoshita06,Kaufman2016,Borgonovi2016}, many-body localization~\cite{Schreiber2015,FanARXIV,WeiARXIV}, quantum correlations~\cite{GarttnerARXIV}, and quantum speed limits~\cite{Mandelstam1945,Margolus1998,MugaBook} consider similar dynamical quantities. They include the survival probability, density imbalance, and out-of-time-ordered correlator (OTOC). Our goal is to characterize the evolution of these quantities at different time scales.

Given the complexity of out-of-equilibrium many-body quantum systems, we take the same approach as Wigner when studying heavy nuclei and use full random matrices (FRM) from the Gaussian orthogonal ensemble (GOE). These are matrices filled with random real numbers and constrained by time-reversal symmetry. The model is unrealistic, as it assumes simultaneous and infinite-range interactions among all particles. But it allows for the derivation of analytical expressions for the observables of interest. 

The analysis of the FRM model assists in the identification of general features and bounds for the evolution of realistic  systems.
The analytical expressions obtained with FRM reveal different behaviors at different time scales. After determining the generic causes of these behaviors, one can propose expressions for the dynamics of realistic chaotic many-body quantum systems.  

We compare the analytical expressions for FRMs with numerical results for the one-dimensional (1D) Heisenberg spin-1/2 model with on-site disorder. This system has been extensively studied in the context of many-body localization~\cite{SantosEscobar2004,Dukesz2009,Nandkishore2015}. It shows a chaotic regime for small disorder~\cite{Avishai2002,Santos2004}, which justifies the comparison with FRMs. The rate of the evolution is faster in the FRM case, but the overall dynamical behavior is similar for both models. 

The basis of our analysis is the survival probability. It gives the probability of finding the initial state later in time and has been investigated since the early days of quantum mechanics~\cite{Khalfin1958}. It is a main quantity in the studies of quantum speed limits~\cite{MugaBook} and decay processes of unstable systems~\cite{Fonda1978}. More recently, it became central to the analysis of localization in noninteracting~\cite{Ketzmerick1992,Huckestein1994} and interacting~\cite{Torres2015,Torres2017} systems. The survival probability is also related~\cite{CampoARXIV} to the analytic continuation of the partition function used to study conformal field theories with holographic duals~\cite{DyerARXIV} and to describe the time behavior of large anti-de Sitter black holes~\cite{Maldacena2001,Papadodimas2015,CotlerARXIV}. 

Our analytical expression for the survival probability for the FRM model covers the entire evolution at all different time scales. Following the same steps for its derivation, we  find analytical expressions for the density imbalance and the OTOC. The density imbalance is measured  in experiments with cold atoms~\cite{Schreiber2015,Bordia2017}. The OTOC~\cite{Maldacena2016} quantifies the degree of noncommutativity in time between two Hermitian operators  that commute at time $t=0$ and has been studied experimentally~\cite{GarttnerARXIV}. Guided by the derivations with FRM, we propose expressions that match very well the numerical evolution of the realistic spin model. 

The short-time dynamics of the survival probability is controlled by the Fourier transform of the envelope of the energy distribution of the initial state, the so-called local density of states (LDOS). When the perturbation that takes the system out of equilibrium is strong, the LDOS is similar to the density of states (DOS). The DOS for the FRM has a semicircle shape, which leads to a decay $\propto {\cal J}_1^2(t)/t^2$, where ${\cal J}_1(t)$ is the Bessel function of first kind~\cite{Torres2014PRA,Torres2014NJP,Torres2014PRAb,Torres2016Entropy}. The initial decay of the density imbalance follows the same behavior, whereas the OTOC goes as ${\cal J}^4_1(t)/t^4$. For the spin system, where only two-body interactions exist, the decay is slower. In this case, maximally spread-out LDOS reach Gaussian shapes~\cite{Torres2014PRA,Torres2014NJP,Torres2014PRAb,Torres2016Entropy,Torres2014PRE,Izrailev2006}, resulting in Gaussian decays. 

The envelope of the oscillations of the term involving the Bessel function decays as $1/t^{3}$ for the survival probability~\cite{TorresKollmar2015,Tavora2016,Tavora2017} and imbalance and as $1/t^{6}$ for the OTOC. These behaviors emerge when the tails of the DOS  fall with the square root of the energy~\cite{Tavora2016,Tavora2017}. In the spin model, the tails of the DOS decay slowly to its energy bounds, which yields smaller power-law exponents.

For long times, but still shorter than the inverse of the mean level spacing (Heisenberg time), the survival probability for both the FRM and the spin model shows a dip below its saturation value, known as correlation hole~\cite{Leviandier1986,Guhr1990,Wilkie1991,Alhassid1992,Gorin2002}. This is an explicit dynamical manifestation of level repulsion in 
systems with discrete spectra~\cite{Torres2017,TorresARXIV}. For yet longer times, the survival probability eventually saturates. Its increase from the bottom of the hole to saturation is nearly linear. We show that the correlation hole appears also for the imbalance and the OTOC.

{\em Hamiltonians and dynamical quantities.}  We consider Hamiltonians $H=H_0 + J V$ that have an unperturbed part $H_0$ and a perturbation $V$ of strength $J$. We set $J=1$  and $\hbar =1$. 

For the 1D spin-1/2 model with onsite disorder,  $L$ sites, and periodic boundary conditions, 
$H_0= \sum_{k=1}^L h_k  S_k^z$ and $V=  \sum_{k=1}^L \vec{S}_k \vec{S}_{k+1}$, where $\vec{S}_k$'s are the spin operators on site $k$. The amplitudes $h_k$ of the static magnetic fields are random numbers from a uniform distribution $[-h,h]$. The total spin in the $z$ direction ${\cal S}^z=\sum_kS_k^z$ is conserved. We study the largest subspace ${\cal S}^z=0$, which has dimension ${\cal N}=L!/(L/2)!^2$. 

When $h=0$ or $h>h_c$, where $h_c$ is the critical point for spatial localization, the eigenvalues can cross, and the level spacing distribution is Poissonian as typical of integrable models. For $0<h<h_c$, the eigenvalues become correlated and repel each other. The level spacing distribution is intermediate between the Wigner-Dyson and the Poissonian distributions. The best agreement with the Wigner-Dyson distribution for ${\cal N} = 12870$ occurs at $h\sim0.5$ \cite{Torres2017}.

In the FRM model, $H_0$ is the diagonal part of the matrix, and $V$ consists of the off-diagonal elements. In the FRM from the GOE, the matrix elements $H_{nm}$ are random numbers from a Gaussian distribution with mean zero. The variance of the elements of $V$ is $\sigma^2$, and for $H_0$, it is $2\sigma^2$. Due to the rotational symmetry, $H_{nm}=H_{mn}=H^{*}_{mn}$ \cite{Guhr1998}. As in the spin model, ${\cal N}$ is the size of the matrix.

The system is initially in one of the eigenstates $| \phi _n \rangle$ of $H_0$. The dynamics starts by switching on the perturbation abruptly. The evolution of the initial state $| \Psi(0) \rangle=| \phi _{n_0} \rangle$ is dictated by $H$, $|\Psi(t) \rangle = e^{-iHt} | \Psi(0) \rangle$. The eigenvalues and eigenstates of $H$  are denoted by $E_{\alpha}$ and $|\psi_{\alpha} \rangle $. The dynamical quantities investigated are listed below.

(i) The survival probability is given by 
\begin{equation}
W_{n_0} (t) =\left| \langle \Psi(0) |  \Psi(t) \rangle \right|^2= \left| \sum_{\alpha}  \left| C_{n_0}^{(\alpha)} \right|^2 e^{-i E_{\alpha}t} \right|^2,
\label{Eq:SP}
\end{equation}
where $C^{(\alpha)}_{n_0}= \langle \psi_{\alpha} |  \Psi(0) \rangle $.

(ii) The imbalance of the spin density for all sites is computed as in \cite{Luitz2016,Lee2017}, 
\begin{equation}
I(t) = \frac{4}{L} \sum_{k=1}^L \langle \Psi(0)| S^z_k(0) S^z_k(t) |  \Psi(0) \rangle .
\label{Eq:Imba} 
\end{equation}

(iii) In terms of spin operators, the OTOC  that we calculate is similar to the one in \cite{FanARXIV},
\begin{equation}
O_{toc}(t) \!=\! \frac{32(L-2)!}{L!{\cal N}}  \! \sum_{n,k,k'} \langle  \phi _n | S^z_{k'} (t) S^z_k (0) S^z_{k'} (t) S^z_k (0) | \phi _n \rangle, 
\label{Eq:OTOC}
\end{equation} 
where we average over all pairs of sites $k'>k$. In the thermal ensemble average, all states $ | \phi _n \rangle$ of the subspace ${\cal N}$ are assumed to contribute equally.

{\em Survival probability.--} We can write Eq.~(\ref{Eq:SP}) in terms of the Fourier transform of  the spectral autocorrelation function as $W_{n_0} (t) =  \int  G(E) e^{-i E t} dE + \overline{W}_{n_0} $, where $G(E)= \sum _{\alpha_1 \neq \alpha_2} |C^{(\alpha_1)}_{n_0} |^2 | C^{(\alpha_2)}_{n_0} |^2 \delta( E - E_{\alpha_1} + E_{\alpha_2}  )  $ and  $ \overline{W}_{n_0} =\sum_{\alpha} |C_{n_0}^{(\alpha)} |^4$ is the infinite time average. 

In the GOE FRM model, the eigenstates are random vectors, so $\langle \overline{W}_{n_0} \rangle_{\text{FRM}} =\overline{W}_{n_0}^{\text{FRM}} = 3/({\cal N}+2)$, where $\langle . \rangle_{\text{FRM}} $ represents the ensemble average. Since the eigenvalues and eigenstates are statistically independent, $G(E)$ is separated into $ \langle \sum _{\alpha_1 \neq \alpha_2} |C^{(\alpha_1)}_{n_0} |^2 | C^{(\alpha_2)}_{n_0} |^2 \rangle_{\text{FRM}} = 1-\overline{W}_{n_0}^{\text{FRM}}  $ and $\langle \delta( E - E_{\alpha_1} + E_{\alpha_2}  ) \rangle_{\text{FRM}} = \int \delta(E - E_{\alpha_1} + E_{\alpha_2}) R_2(E_{\alpha_1},E_{\alpha_2})dE_{\alpha_1} dE_{\alpha_2}/[{\cal N}({\cal N}-1)]$, where $R_2(E_{\alpha_1},E_{\alpha_2})$ is the two-point correlation function. $R_2$ splits in the one-point correlation function, which is simply the DOS, and the two-level cluster function~\cite{MehtaBook}. As ${\cal N} \rightarrow \infty$, the DOS converges to the Wigner semicircle law $$ \rho(E)= \frac{2 {\cal N}}{\pi \varepsilon} \sqrt{1 - \left(\frac{E}{\varepsilon}\right)^2},$$ where $2\varepsilon$ is the length of the spectrum. 

The Fourier transform of the semicircle leads to a term $\propto {\cal J}_1(\varepsilon t)/t$ \cite{Torres2014PRA}. The Fourier transform of the two-level cluster function gives the two-level form factor $b_2(Dt/2\pi)$, where $D$ is the mean level spacing~\cite{MehtaBook,Guhr1998}. In the large ${\cal N}$ limit, $D \approx 1/\rho(0)$. Therefore,
\begin{equation}
W_{n_0}^{\text{FRM}} (t) \!=\!  \frac{1-\overline{W}_{n_0}^{\text{FRM}}  }{{\cal N} -1} \left[ 4 {\cal N} \cfrac{{\cal J}_1^2 (\varepsilon t)}{(\varepsilon t)^2}
- b_2 \left( \cfrac{\varepsilon t}{4 {\cal N} } \right) \right] + \overline{W}_{n_0}^{\text{FRM}}  ,
\label{Eq:Wo}
\end{equation}
where  $b_2(\overline{t}) = [1-2\overline{t} + \overline{t} \ln(1+2 \overline{t})] \Theta (1- \overline{t}) + \{-1 + \overline{t} \ln [ (2 \overline{t}+1)/(2 \overline{t} -1) ] \} \Theta(\overline{t}-1)$ and $\Theta$ is the Heaviside step function. 

\begin{figure}[ht!]
\hspace{-0.6cm}
\includegraphics*[width=3.5in]{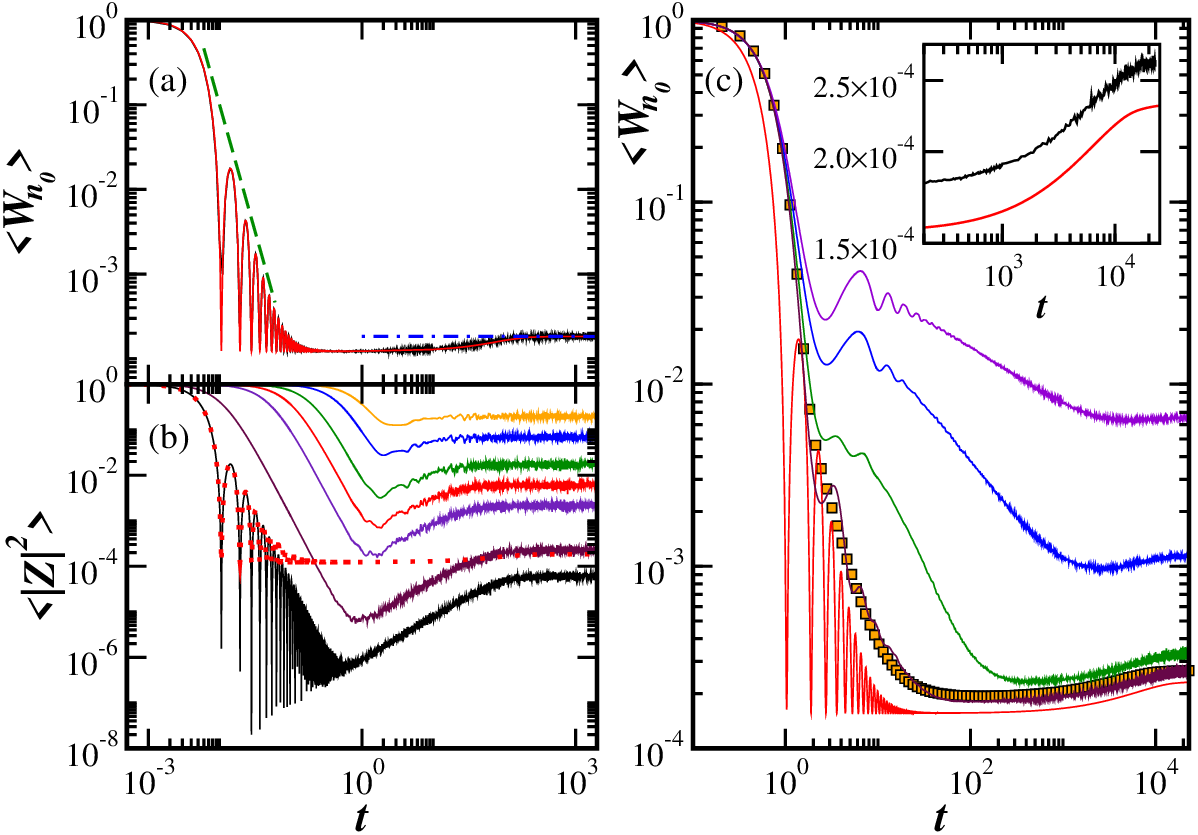}
\caption{Survival probability and $|Z(\beta + it)|^2$. In (a), GOE FRM.  Numerical results and Eq.~(\ref{Eq:Wo}) are superposed; $1/t^{3}$ decay (the dashed curve), saturation value (the dot-dashed curve). In (b), the solid lines from bottom to top give $|Z(\beta + it)|^2$ with $\beta=0,0.01,0.05,0.1, 0.2, 0.5, 1$; the dotted curve is Eq.~(\ref{Eq:Wo}). In (c), the solid lines from bottom to top are as follows: Eq.~(\ref{Eq:Wo}) and numerical results for the spin model with $h=0.5, 1, 1.5, 2$. The squares correspond to the fitting curve for $h=0.5$. The FRM is rescaled, so the DOS of both models have the same width. The inset of (c): Eq.~(\ref{Eq:Wo}) (bottom) and time average for $h=0.5$ (top). In (a) and (b), the averages over 200 disorder realizations; ${\cal N}=16\,384$, $\sigma^2=2$. In (c), the average over $10^5$ data; ${\cal N}=12\,870$.}
\label{Fig01}
\end{figure}

In Fig.~\ref{Fig01} (a), we compare Eq.~(\ref{Eq:Wo}) with the numerical results for the GOE FRM. The agreement is excellent; the two curves can hardly be distinguished. 

The initial evolution of $W_{n_0}^{\text{FRM}} (t) $ is controlled by the term with the Bessel function, which leads to oscillations that decay as $1/t^{3}$, as indicated by the dashed line in Fig.~\ref{Fig01} (a). 
The correlation hole, corresponding to the full time interval where $W_{n_0}^{\text{FRM}} (t)$ is below $\overline{W}_{n_0}^{\text{FRM}}$, is caused by $b_2(\overline{t})$. As we approach the Heisenberg time, the hole fades away, and  the dynamics eventually saturates at $\overline{W}_{n_0}^{\text{FRM}}$.

The correlation hole is a direct probe of short- and long-range correlations between the eigenvalues. For level statistics given by the Poissonian distribution, $b_2(\overline{t})=0$, and the hole is nonexistent. 

In Fig.~\ref{Fig01} (b), we compare Eq.~(\ref{Eq:Wo}) (the dotted line) with numerical results for the analytic continuation of the partition function, $|Z(\beta + it)|^2 = \sum_{\alpha} \exp[-(\beta + it) E_{\alpha}]/Z(\beta)$ (the solid lines). As discussed in Ref.~\cite{CampoARXIV}, $|Z(\beta + it)|^2$ is analogous to the survival probability if one considers as initial state, a thermofield-double state, that is $|\Psi(0) \rangle = \sum_{\alpha} \exp(-\beta E_{\alpha}/2) |\psi_{\alpha} \rangle /\sqrt{Z(\beta)}$. As illustrated in Fig.~\ref{Fig01} (b), the results for $|Z(\beta + it)|^2$  for GOE FRM show qualitative agreement with $W_{n_0}^{\text{FRM}} (t)$. The survival probability and $|Z(\beta + it)|^2$ for $\beta=0$ decay initially as $ {\cal J}_1^2 (\varepsilon t)/(\varepsilon t)^2$, and all curves in Fig.~\ref{Fig01} (b) show correlation holes. However, this comparison has limitations, since in quench dynamics $C^{(\alpha)}_{n_0}$ cannot be chosen independently of $H_0$ and $H$ as performed for the thermofield state. Contrary to  $|Z(\beta + it)|^2$, $W_{n_0}(t)$ depends on the quench protocol. 

Figure~\ref{Fig01} (c) depicts the survival probability for the spin model with different disorder strengths. The curves are averages over disorder realizations and $0.1{\cal N}$ initial states with energy in the middle of the spectrum. Even deep in the chaotic regime ($h=0.5$), the decay of $\langle W_{n_0} (t) \rangle$ is slower than that for the FRM model, being bounded by Eq.~(\ref{Eq:Wo}). This is caused by two related factors typical of realistic systems with two-body interactions:  the Gaussian shape of the DOS~\cite{Brody1981} and the lack of full ergodicity of the eigenstates.

Using as a reference the steps for the analytical derivation of $G(E)$ for FRM, namely that the $R_2$ function splits into the DOS and the two-level cluster function, we look for an expression that can reproduce the evolution of the chaotic spin model. We take into account the following features of the realistic system: (i) The Fourier transform of a Gaussian LDOS gives a Gaussian decay at short times $e^{-w^2 t^2} $, where $w$ is the width of the energy distribution~\cite{Torres2014PRA,Torres2014NJP,Torres2014PRAb,Torres2014PRE,Izrailev2006}, (ii) this distribution is bounded in energy~\cite{Tavora2016,Tavora2017} and nearly constant at the edges, which causes a power-law behavior $\propto 1/t^{2}$, and
(iii) the presence of level repulsion induces the correlation hole at long times. These aspects, together with the saturation of  $\langle W_{n_0} (t) \rangle$, motivate the expression 
\begin{equation}\label{eq:SPD}
\langle W_{n_0} (t) \rangle \!=\!  \frac{1- \langle \overline{W}_{n_0} \rangle}{{\cal N} -1} \left[ {\cal N} \frac{g(t)}{g(0)} - b_2 \left(\frac{w t}{ {\cal N}}\right) \right] + \langle \overline{W}_{n_0} \rangle\,,
\end{equation}
where $g(t) = e^{-w^2 t^2} + A (1-e^{-w^2 t^2} )/(w^2 t^2)$ and $A$ is a fitting constant. Apart from the first term, which depends on the shape and tails of the energy distribution, Eq.\eqref{eq:SPD} is equal to Eq.(\ref{Eq:Wo}). It is impressive that, with a single fitting constant, our expression captures so well the entire evolution of $\langle W_{n_0} (t) \rangle$ for $h=0.5$ as seen in Fig.~\ref{Fig01} (c).

The inset of Fig.~\ref{Fig01} (c) confirms that $b_2$ is the appropriate function to describe the correlation hole also for the chaotic spin system. The $h=0.5$ curve follows closely the FRM analytical expression. This indicates that the long-time behavior of realistic chaotic many-body systems (before saturation) depends only on the correlations in the eigenvalues, not on details of the model, such as the shape of the DOS and structure of the eigenstates.

The origin of the $1/t^3$ decay for the FRM model is the square-root edge of the DOS. This power-law exponent is observed also for the Sachdev-Ye-Kitaev model~\cite{CotlerARXIV,BagretsARXIV} where the DOS is also a semicircle at the edges~\cite{Garcia2016,GarciaARXIV} and for ($1+1$)-dimensional conformal field theories with a gravity dual~\cite{DyerARXIV}. Since field theories with holographic duals set bounds to certain dynamical coefficients~\cite{Kovtun2005}, one may speculate whether the $1/t^3$ behavior is a general bound to the decay of the survival probability and related quantities of generic lattice many-body quantum systems.  If we replace the Gaussian distribution of the random entries of the FRM by distributions involving higher even powers, it is possible to achieve DOS whose tails go as $|E-E_0|^\xi$ where $\xi=3/2, 5/2, \ldots$ and $E_0$ is the edge of the spectrum~\cite{Brezin1978}, which would lead to decays faster than $1/t^3$. Whether there may be realistic systems with such DOS is an open question.

{\em Density imbalance.} Level repulsion manifests itself not only as the correlation hole of the survival probability. It is revealed also in the long-time evolution of experimental observables such as the spin density imbalance. 

\begin{figure}[ht!]
\hspace{-0.6cm}
\centering
\includegraphics*[width=3.5in]{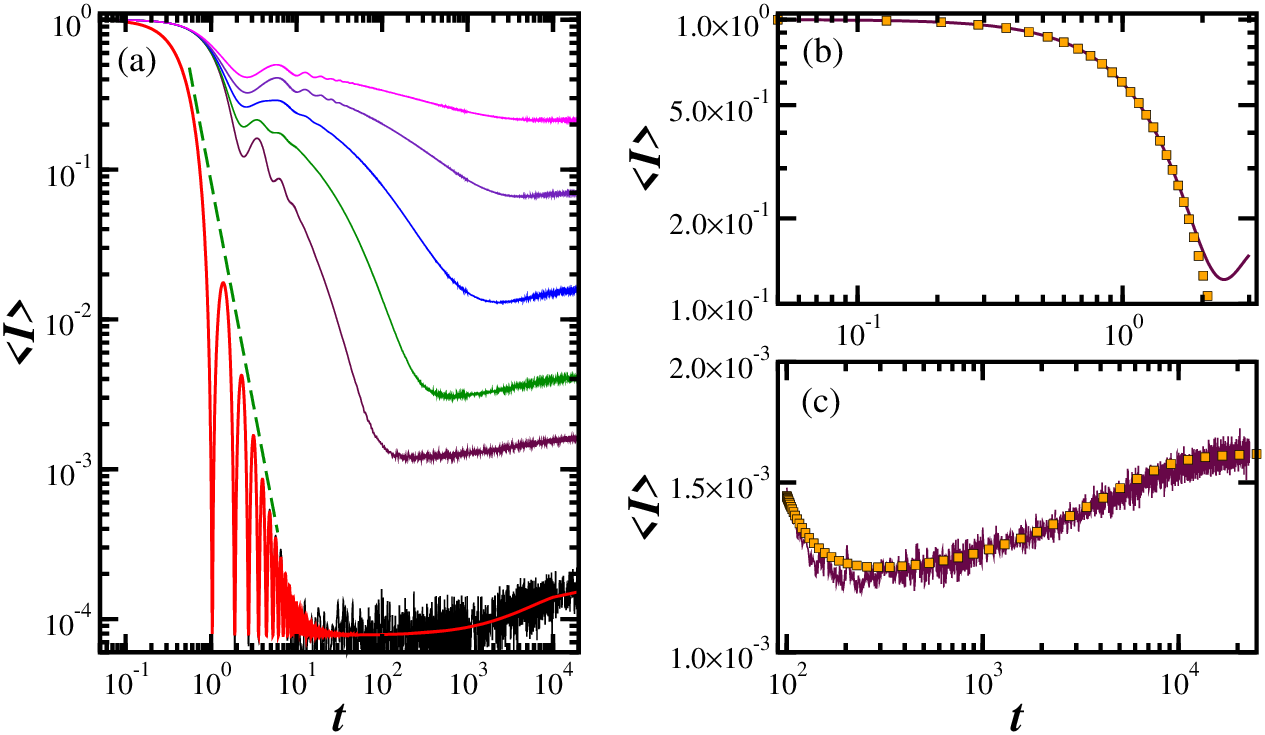}
\caption{Density imbalance for the FRM and the spin model. In (a) from bottom to top, the FRM  (numerical and analytical curves) and disorder strength $h=0.5,  1, 1.5, 2, 2.5$; $1/t^3$ (the dashed curve). In (b) and (c), the numerical result (the solid curve) and fitting (the squares) for $h=0.5$. In (b), the short-time dynamics with Gaussian behavior. In (c), the long-time evolution  fitted with a power-law decay and the $b_2(\overline{t})$ function. Averages over $10^4$ random realizations; ${\cal N}=12\,870$.}
\label{Fig02}
\end{figure}

The curves for the density imbalance for the FRM model and for the disordered spin system with different values of $h$ show a dip below the saturation value as illustrated in Fig.~\ref{Fig02} (a). As $h$ increases above $0.5$ and the realistic system moves away from the chaotic region, the hole in Fig.~\ref{Fig02} (a) shows the same features of the hole in Fig.~\ref{Fig01} (c). It gets less deep, its time interval shrinks, and the moment when it first appears gets deferred to longer times. This is consistent with the fact that the long-range correlations in the eigenvalues diminish as the realistic system moves towards a localized phase. The depth of the correlation hole has been used to signal the metal-insulator transition in Refs. \cite{Torres2017,TorresARXIV}.

To obtain an analytical expression for the density imbalance, we refer to the equation  $O(t)=  \int K(E) e^{-iEt} dE + \overline{O} $ for a general observable $O$, where $K(E) =  \sum _{\alpha_1 \neq \alpha_2} C^{(\alpha_1)}_{n_0}  C^{(\alpha_2)}_{n_0} O_{\alpha_1 \alpha_2} \delta(E - E_{\alpha_1} + E_{\alpha_2})$ with $O_{\alpha_1 \alpha_2} =\langle \psi_{\alpha_1} | O | \psi_{\alpha_2} \rangle$ and $\overline{O} = \sum _{\alpha}  |C^{(\alpha)}_{n_0}|^2 O_{\alpha \alpha}$  is the infinite time average. In the FRM model, where the eigenvalues, eigenstates, and $O_{\alpha_1 \alpha_2}$ are statistically independent, we can separate $K(E)$ into $\langle  \sum _{\alpha_1 \neq \alpha_2} C^{(\alpha_1)}_{n_0}  C^{(\alpha_2)}_{n_0} O_{\alpha_1 \alpha_2} \rangle_{\text{FRM}} = O(0) - \overline{O}^{\text{FRM}}$ and $\langle  \delta(E - E_{\alpha_1} + E_{\alpha_2}) \rangle_{\text{FRM}}$, already computed for Eq.~(\ref{Eq:Wo}). 

Using the reasoning above, we obtain the following expression for the density imbalance:
\begin{equation}
I^{\text{FRM}} (t) \!=\! \frac{I(0)-\overline{I}^{\text{FRM}}}{{\cal N} - 1} \left[ 4 {\cal N} \cfrac{{\cal J}_1^2 (\varepsilon t)}{(\varepsilon t)^2}
 - b_2 \left( \cfrac{\varepsilon t}{4 {\cal N} } \right)\right]  + \overline{I}^{\text{FRM}} ,
 \label{Eq:Oo}
\end{equation}
where $\overline{I}^{\text{FRM}}  = 2 I(0)/({\cal N} +2)$. The result is very similar to that for the survival probability, leading also to the $1/t^{3}$ decay of the oscillations as seen in Fig.~\ref{Fig02} (a).

The decay of the density imbalance for the spin model is bounded by Eq.~(\ref{Eq:Oo}). It shows a power-law behavior also in the chaotic domain, which indicates that algebraic decays  are not exclusive to systems in the vicinity of a localized phase.

The relaxation of $I(t)$ for the disordered spin model was investigated in Ref. \cite{Luitz2016}. There, a fitting function with nine free parameters was proposed for the intermediate times where the power-law behavior is observed. We add to this picture the description of the short- and long-time dynamics. 

The  imbalance for the spin system follows closely what happens for the survival probability. The initial decay, up to $wt\sim2$, is Gaussian as shown in Fig.~\ref{Fig02} (b).

The correlation hole emerges at long times and is shown in Fig.~\ref{Fig02} (c). The numerical curve for $h=0.5$ is fitted with the function $A t^{-B} - C b_2 \left(\frac{w t}{ {\cal N}}\right)$, where $A$, $B$, and $C$ are fitting constants. We use the same $b_2 (\overline{t})$ used for the survival probability in Fig.~\ref{Fig01} (c). The agreement is extremely good, covering a large time interval all the way to saturation.

{\em Out-of-time-ordered correlator.} Analogous to what happens for the density imbalance, the evolution of the OTOC for the FRM model is initially very fast and later shows oscillations that decay as $1/t^6$. The OTOC involves the four-point correlation function $R_4 (E_{\alpha_1} , E_{\alpha_2} , E_{\alpha_3}, E_{\alpha_4})$ derived from the ensemble average $\langle \delta (E - E_{\alpha_1} + E_{\alpha_2} - E_{\alpha_3} + E_{\alpha_4} ) \rangle_{\text{FRM}}$. $R_4$ can be expressed as the determinant of a single spectral kernel which is known explicitly~\cite{MehtaBook}. For short and intermediate times, the leading contribution to the Fourier transform of $R_4$ is proportional to  ${\cal J}_1^4 (\varepsilon t)/(\varepsilon t)^4$, which causes the $1/t^6$ decay. At long times, $b_2^2(Dt/2\pi)$ becomes dominant and causes the correlation hole.

\begin{figure}[h!]
\includegraphics*[width=3.5in]{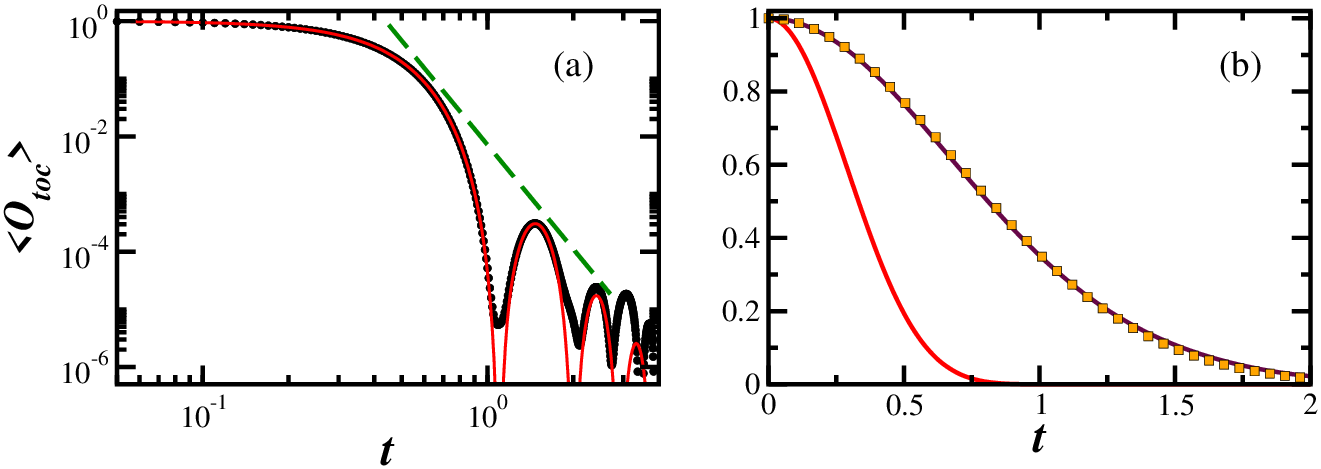}
\caption{OTOC for FRM (a) and compared with the disordered model for $h=0.5$ (b). In (a), ${\cal J}_1^4 (\varepsilon t)/(\varepsilon t)^4$ (the solid curve), numerical results (the circles), and $1/t^6$ (the dashed curve). In (b), the FRM (bottom) and $h=0.5$ (top); numerical curve (the solid curve) and Gaussian fit (the squares). Averages over 340 (FRM) and 100 (spin model) disorder realizations; ${\cal N}=3432$.}
\label{Fig03}
\end{figure}

The $1/t^6$ behavior of the OTOC is shown in Fig.~\ref{Fig03} (a). The agreement between the numerical data and the analytical prediction from the FRM is very good. In Fig.~\ref{Fig03} (b), the analytical curve for the FRM model is compared with the decay for the disordered spin system with $h=0.5$. The decay of the latter is slower and exhibits a Gaussian behavior  for short times.

The survival probability, and therefore $I(t)$ and  $O_{toc}(t)$, are not self-averaging~\cite{Prange1997}. The size of the ensemble of random matrices needed to reasonably expose the correlation hole for the density imbalance and the OTOC is significantly larger than for $\langle W_{n_0}(t)\rangle$.

{\em Conclusion.}  We have found analytical expressions for the evolution of the survival probability, density imbalance, and OTOC for a FRM model. These observables are central to theoretical and experimental studies of quantum systems out of equilibrium. The analytical findings were compared with numerical results for a 1D-disordered spin-1/2 system. The power-law decays, for intermediate times, and dips below the saturation values, for longer times, revealed by the FRM model appeared also for the chaotic spin model. The identification of these generic properties helped us finding and justifying functions that describe very well the numerical evolution of the spin model at different time scales. This approach can be used also for describing equivalent realistic lattice many-body quantum systems with level repulsion.

{\em Acknowledgments.} E.J.T.-H acknowledges funding from CONACyT and VIEP-BUAP, Mexico.  He is also grateful to LNS-BUAP for allowing the use of their supercomputing facility. A.M.G.-G acknowledges partial financial support from a 
QuantEmX grant from ICAM and the Gordon and Betty Moore Foundation through Grant No. GBMF5305. L.F.S. was supported by the NSF Grant No. DMR-1603418.


\end{document}